\documentclass[10pt]{article}

\usepackage[T1]{fontenc}
\usepackage[left=2cm, right=2cm, top = 2.5cm, bottom = 4cm]{geometry}
\usepackage{amsmath}
\usepackage{amssymb}
\usepackage{braket}
\usepackage{graphicx}
\usepackage[sorting=none]{biblatex}
\usepackage{hyperref}
\hypersetup{
    colorlinks = true,
    linkcolor = blue,
    anchorcolor = blue,
    citecolor = blue,
    filecolor = blue,
    urlcolor = blue
}

\usepackage{xcolor}
\usepackage{fancyhdr}
\newcommand{\op}[1]{\hat{#1}}
\newcommand{\mi}{\mathrm{i}}
\newcommand{\lpar}[1]{\left(#1\right)}
\newcommand{\lspar}[1]{\left[#1\right]}

\newcommand{\tr}{\mathrm{tr}}

\title{Moving mirror-field dynamics under intrinsic decoherence}
\author{\Large{Alejandro R. Urz\'ua$^{1}$ and H\'ector M. Moya-Cessa$^{2}$}\\ \\ 
$^{1}$Instituto de Ciencias F\'isicas, Cuernavaca, Morelos\\
$^{2}$Instituto Nacional de Astrof\'isica, \'Optica y Electr\'onica, Tonantzintla, Puebla}

\begin{document}
\maketitle
\thispagestyle{fancy}

\begin{abstract}
We study the decaying dynamics in the mirror-field interaction by means of the intrinsic decoherence scheme. Factorization of the mirror-field Hamiltonian with the use of displacement operators, allows us to calculate the explicit solution to Milburn's equation for arbitrary initial conditions. We show  expectation values,  correlations, and Husimi functions for the solutions obtained.
\end{abstract}

\section{Introduction}\label{S1:intro}
The presence of decoherence in quantum mechanical systems is an undesirable effect coming mainly from two sources: interaction between internal degrees of freedom, and the interaction with the environment via some coupling \cite{Bowen2015}. Intrinsic decoherence \cite{Milburn1991, Rajagopal1996} is a  topic that has recently raised great interest, as it has become a way to describe phase decay which is one of the most damaging effects of quantum physical systems, making mandatory its study. Therefore, quantifying the rate at which an isolated system loses its phase coherence in time is needed in order to try to avoid such a harmful phenomenon. We will apply its study here to an optomechanical system. Such systems deal with the interaction of light and mechanical devices coupled by means of their internal degrees of freedom in the most simple setting \cite{Marquardt2009, Aspelmeyer2014, Milburn2011}. One realization of such systems is a cavity containing a quantized electromagnetic field coupled to a mirror experiencing harmonic motion; in which interesting phenomena can occur due to the feedback action between the subsystems, like optomechanical entanglement \cite{Vitali2007}, or the creation of real photons in the cavity due to the dynamical Casimir effect and how the phase damping of the intrinsic decoherence intervene in the time dynamics \cite{Schtzhold2005}.

Since the seminal work by Milburn \cite{Milburn1991} on the possibility to model  decoherence based on a simple modification of unitary Schrödinger evolution, much work has been done pointing in several directions, depending on how  decoherence is tried to be diminished, or toked in advance. Moreover, intrinsic decoherence produces a Master Equation of the Lindblad type making it possible to  compare with other methods that describe  open systems, like master equations and quantum trajectories.

The last decade has been very prolific in this field we want to board, the community has taken the task to study a vast amount of systems that can experience intrinsic decoherence. We can see research on quantification of non-classicality \cite{Benabdallah2022}, qubits for quantum information processing \cite{Dahbi2023}, bipolar spin systems \cite{Oumennana2022}, quantum dot correlations \cite{Chaouki2022}, Heisenberg XYZ spin chains, quantum-memory-assisted two-qubit, and the temporal evolution of quantum correlations \cite{Essakhi2022, AitChlih2022, AitChlih2021}, quantum-memory-assisted entropic uncertainty, mixedness, and entanglement dynamics in two-qubit system \cite{Mohamed2022}, symmetric spin-orbit model \cite{Mohamed2022b}, two-qubit quantum Fisher information \cite{Alenezi2022}, two-qubit maximally entangled Bell states \cite{Mohamed2022c}, trapped ions \cite{Mohamed2022d}, isolated Heisenber and Aubry-André spin models \cite{Wu2017}, two coupled quantum dots \cite{Mirzaei2022}, N-level atomic system \cite{Anwar2019}, two-coupled qubit two-level cavity \cite{Mohamed2021}, two-level atom \cite{Anwar2018}, state transfer in spin channels \cite{Hu2009}, Heisenber anisotropic interaction \cite{Muthuganesan2021}, nonlocal advantage of quantum coherence \cite{Li2021}, qutrit teleportation \cite{Naderi2019}, and quantum dense coding \cite{Zhang2016}. When we look at specific examples related optomechanical systems, we find notable works on Jaynes-Cummings under intrinsic decoherence \cite{hessian1999, Kuang1995, Obada2004}, bimodal multiquanta Jaynes-Cummings \cite{Obada1998,
}, entanglement two Tavis-Cummings (no-RWA JC) \cite{Guo2009}, and ultra-strong coupled harmonic oscillator in cavities \cite{Habarrih2023}.

In this manuscript, we analyze the effects of intrinsic decoherence in a mirror and a quantized field coupled by means of radiation pressure and harmonic mechanical motion. By using solutions recently proposed by us to study the one-dimensional displaced harmonic oscillator \cite{Urzua2022} and the three-coupled one-dimensional harmonic oscillators \cite{Urzua2023}, we solve the complete form of the equation instead of the master equations Milburn obtained at first.

\section{Moving mirror-field interaction}\label{S2:mfield}
The standard setup for the moving mirror-field interaction is a cavity enclosing the radiation, such as the quintessential Fabry-P\'erot interferometer \cite{Vaughan2017}, where two perfectly reflecting mirrors, one fixed and the other experimenting harmonic motion described by the macroscopic canonical position $\op{q}(t)$; inside the cavity, there's a coherent driven field $\op{a}\lpar{\op{a}^{\dagger}}$ pumped by a tunable laser \cite{Aspelmeyer2014}. The interaction between the mirror and the field is given by radiation pressure coupled with the (microscopical) number of modes of the field and the (macroscopical) position quadrature of the mechanical device. The Hamiltonian for the moving mirror-field interaction then reads \cite{Bose1997},
\begin{eqnarray}\label{hinit}
     \op{H} = \omega\op{n} + \nu\op{N} + \chi\op{n}\lpar{\op{b}^{\dagger} + \op{b}}
\end{eqnarray}
where we have defined the number operators: $\op{n} = \op{a}^{\dagger}\op{a}$ for the field, and $\op{N} = \op{b}^{\dagger}\op{b}$ for the mirror. The parameters $\omega$, $\nu$, and $\chi$ are the frequencies of the field, the mirror, and the coupling strength, respectively. This Hamiltonian may be rewritten in a diagonal form with the help of a composite \emph{displacement operator} in the mirror eigenbasis $(\op{b})$ as
\begin{eqnarray}\label{hdiag}
    \op{H} = \op{D}^{\dagger}\lpar{\frac{\chi\op{n}}{\nu}}\lspar{\nu\op{N} + \omega\op{n} - \frac{\chi^{2}}{\nu^{2}}\op{n}^{2}}\op{D}\lpar{\frac{\chi \op{n}}{\nu}},
\end{eqnarray}
where $\op{D}^{\dagger}\lpar{\tfrac{\chi\op{n}}{\nu}} := \exp\lpar{\lspar{\tfrac{\chi}{\nu}\op{b}^{\dagger} - \tfrac{\chi^{*}}{\nu^{*}}\op{b}}\otimes\op{n}}$, that is a tensor element of the two eigenbases, acting as a true displacement operator on the mirror basis, and as an exponential of the number operator in the field basis.\\

\begin{figure}[htbp]
    \centering
    \includegraphics[width = \columnwidth]{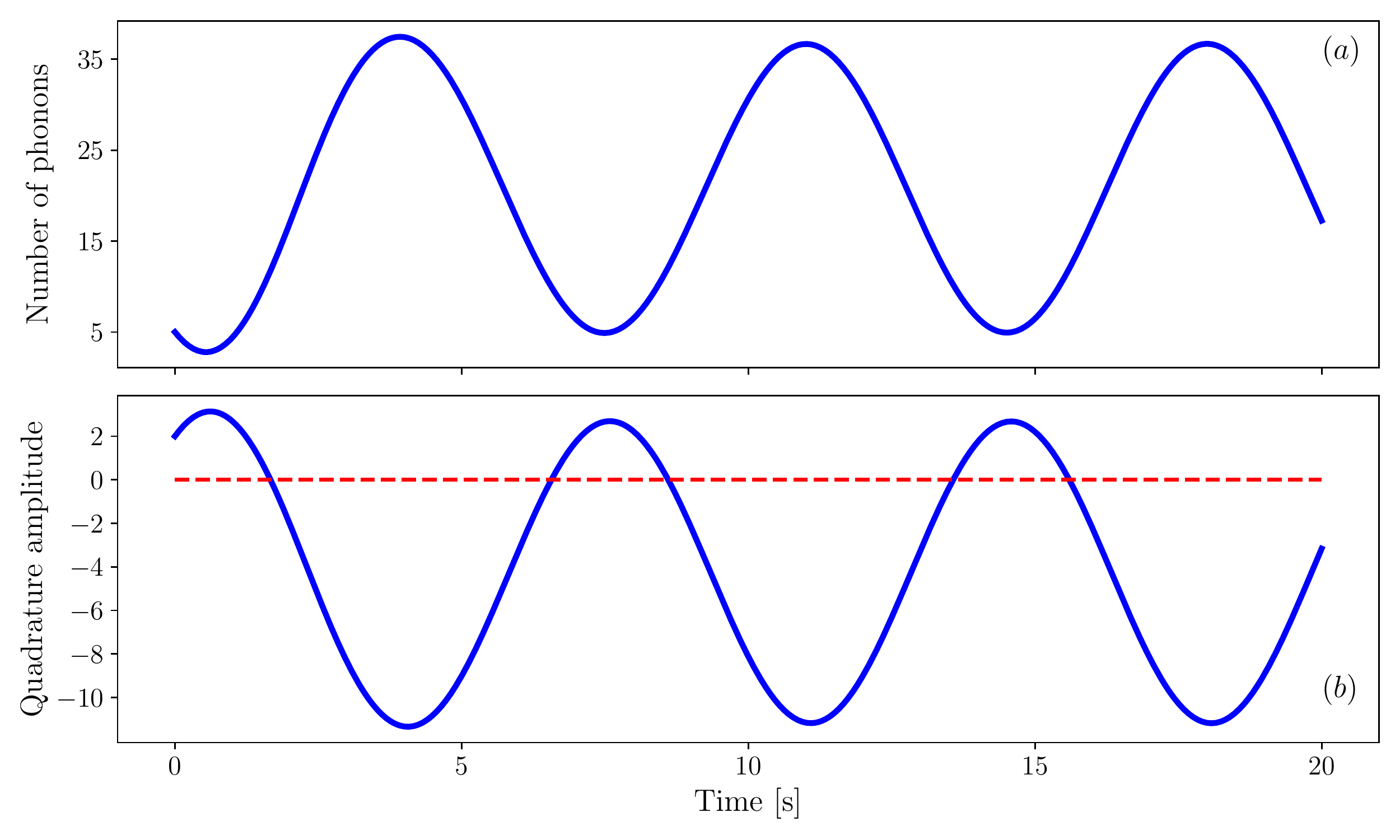}
    \caption{Expectation values for $(a)$ the number of phonon modes $\braket{\op{N}}$ and $(b)$ the position quadrature $\braket{\op{q}}$ in the mirror, using \eqref{ansol}.}
    \label{fig:schr_sol}
\end{figure}

\paragraph{Analytical solution} There exists a straightforward solution of the Schr\"odinger equation associated to Hamiltonian \eqref{hdiag} when the initial condition is set to $\ket{\alpha}_{f}\otimes\ket{\beta}_{m}$, giving the expression  (see refs. \cite{Bose1997, Mancini1997} for the detailed solution via exponential disentanglement),
\begin{equation}\label{ansol}
    \ket{\psi(t)} = e^{-\vert\alpha\vert^{2}/2}\sum\limits_{n = 0}^{\infty}\frac{\alpha^{n}}{\sqrt{n!}}e^{\mi (\chi/\nu)^{2} n^{2}\lpar{t - \sin t}}\ket{n}\ket{\phi_n}
\end{equation}
where $\ket{\phi_n}\equiv \ket{\beta e^{-\mi t} + (\chi/\nu)n(1-e^{-\mi t})}$; with $\beta$ the amplitude of the initial coherent  state of the mirror; $\alpha$ the amplitude of the initial coherent state of the field and $\ket{n}$ are number states of the quantized field. This solution shows a constant number of photon modes in the field, and periodic amplitudes in the phonon modes and position (quadrature) in the mirror, see Figure \ref{fig:schr_sol}. We expect that the behavior in the field is preserved after the decoherence drive is applied and the amplitudes and position in the mirror might experience decaying modulation depending on the parameters involved in the intrinsic decoherence dynamic.

\paragraph{Intrinsic decoherence scheme}
According to the seminal work of Milburn \cite{Milburn1991}, we can observe decaying dynamics in the Schr\"odinger evolution of quantum systems modifying the mathematical structure of the governing equation. This modification takes into account the disorder-induced stochastic phase changes in the Hamiltonian of the system at very short times. In its closed form, the density matrix evolution turns out to be
\begin{equation}\label{irho}
    \dot{\rho}_{S}\mapsto\dot{\rho}_{M}=\gamma \lpar{e^{-\mi\frac{\op{H}}{\gamma}}\op{\rho} e^{\mi\frac{\op{H}}{\gamma}}- \op{\rho}},
\end{equation}
where $\op{H}$ is the system's Hamiltonian, and $\gamma$ is the decay rate parameter of the intrinsic decoherence, giving the attributes to determine the time scale of the coherence suppression. We can, of course, expand equation \eqref{irho} in terms of $\gamma^{-1}$ to obtain the $n$-th order of approximation, which leads to the emergence of a Lindblad master equation. But on the other side, the equation \eqref{irho} can be solved analytically using the adequate \emph{ansatz}
\begin{equation}\label{rhoin}
    \op{\rho}(t)=e^{-\gamma t}e^{\op{S}t}\op{\rho}(0),
\end{equation}
where we have defined the \emph{superoperator} \cite{MoyaCessa2006}
\begin{equation*}
    \op{S}\op{\rho}=\gamma e^{-\mi\frac{\op{H}}{\gamma}}\op{\rho} e^{\mi\frac{\op{H}}{\gamma}},
\end{equation*}
such that the Taylor expansion of the action gives
\begin{equation}\label{rhotay}
    e^{\op{S}t}\op{\rho}(0)=\sum_{k=0}^{\infty}\frac{(\gamma t)^k}{k!}\, \op{\rho}_{k},
\end{equation}
with the $k$-th element of the density matrix $ \op{\rho}$ defined by
\begin{equation}\label{rhok}
    \op{\rho}_{k} := \ket{\psi_{k}}\bra{\psi_{k}},\qquad \ket{\psi_{k}} = e^{-\frac{\mi k}{\gamma}\op{H}}\ket{\psi(0)}.
\end{equation}
Then, using \eqref{rhok}, inserting in \eqref{rhotay}, and putting all together with \eqref{rhoin}, we arrive at the explicit form of the solution, given by
\begin{equation}\label{rhot}
    \op{\rho}(t) = e^{-\gamma t}\sum\limits_{k = 0}^{\infty}\frac{\lpar{\gamma t}^{k}}{k!} \ket{\psi_{k}}\bra{\psi_{k}}.
\end{equation}

For the particular system we are dealing with, the $k$-th element of the wavefunction takes the form
\begin{equation}\label{psik}
    \ket{\psi_{k}} = \op{D}^{\dagger}\lpar{\frac{\chi\op{n}}{\nu}}e^{-\frac{\mi k}{\gamma}\lspar{\nu\op{N} + \omega\op{n} - \frac{\chi^{2}}{\nu}\op{n}^{2}}}\op{D}\lpar{\frac{\chi \op{n}}{\nu}}\ket{\psi(0)},
\end{equation}
that recalls the analytical Schr\"odinger solution. At this point, we can apply \eqref{psik} to a direct product of initial wavefunctions for the mirror and the field: $\ket{\psi(0)} = \ket{\phi(0)}_{m}\otimes\ket{\theta(0)}_{f}$.\\

This depicted scheme is a mathematical procedure relying on a \emph{Poisson model} for the short-time stochastic behavior of the system, as Milburn stated. On the other side, selecting suitable initial conditions $\ket{\psi(0)}$ is dependent on the particular system we have, in the case of optomechanical devices that coupled electromagnetic fields and harmonic moving mirror, the selection of coherent states $\ket{\alpha}$ for the field and $\ket{\beta}$ for the mirror can be viewed as obvious ---in broad sense we have, of course, a system of two-coupled harmonic oscillators.--- Finally, the existence of the limit in \eqref{rhot} precludes how many analytical solutions can we obtain, since for some initial wavefunctions, the limit may not exist, nevertheless it can be always solved numerically at some truncation term $<\infty$.

\section{Solution and observables}
\subsection{Expectation values}
If we have a suitable operator $\op{A}$ that represents an observable $a$, the expectation value of this operator can be obtained from the density matrix $\op{\rho}$ as $\tr\lpar{\op{\rho} \op{A}} \equiv \bra{\psi}\op{A}\ket{\psi}$, for some wavefunction $\ket{\psi}$ in the eigenbasis of the system \cite{Sakurai2020}. Now, for the density matrix in \eqref{rhot} that depends on the $k$-th component $\op{\rho}_{k}$, the calculation of the expectation value reduces to
\begin{equation}\label{expv}
    \langle\op{A}\rangle := e^{-\gamma t}\sum\limits_{k = 0}^{\infty}\frac{\lpar{\gamma t}^{k}}{k!}\bra{\psi_{k}}\op{A}\ket{\psi_{k}},
\end{equation}
for $\ket{\psi_{k}}$ defined by \eqref{psik}. Thus, given the form of $\ket{\psi_{k}}$, we need to obtain the action of the displacement operators and the exponential maps onto the operator $\op{A}$. At this point, we are interested in the most descriptive features in the optomechanical system: the number of photon modes in the field, that we predict has to be constant; the number of mechanical modes and quadrature in the mirror, that should have a decaying dynamic depending on the decoherence rate $\gamma$; a set of statistical estimators, like the Hong-ou-Mandel parameters, and the covariance; finally, a representation in phase-space of the dynamics via the Husimi $\op{Q}$-function.

\subsubsection{Number of phonon modes $\braket{\op{N}}$ in the mirror}
We start calculating the expectation value of the number of phonon modes in the mirror. The left-right action of the wavefunction $\ket{\psi_{k}}$ onto the operator $\op{N}$ is straightforward, since we know the expansion
\begin{equation}
\begin{aligned}
    \bra{\psi_{k}}\op{N}\ket{\psi_{k}} &= \bra{\psi(0)}\lpar{\op{D}^{\dagger}\lpar{\frac{\chi\op{n}}{\nu}}e^{\frac{\mi k}{\gamma}\lspar{\nu\op{N} + \omega\op{n} - \frac{\chi^{2}}{\nu}\op{n}^{2}}}\op{D}\lpar{\frac{\chi \op{n}}{\nu}}}\op{N}
    \lpar{\op{D}^{\dagger}\lpar{\frac{\chi\op{n}}{\nu}}e^{-\frac{\mi k}{\gamma}\lspar{\nu\op{N} + \omega\op{n} - \frac{\chi^{2}}{\nu}\op{n}^{2}}}\op{D}\lpar{\frac{\chi \op{n}}{\nu}}}\ket{\psi(0)}\\
    &= \bra{\psi(0)}\lpar{\op{D}^{\dagger}\lpar{\frac{\chi\op{n}}{\nu}}e^{\frac{\mi k}{\gamma}\lspar{\nu\op{N} + \omega\op{n} - \frac{\chi^{2}}{\nu}\op{n}^{2}}}}\times\lpar{\op{N}-\frac{\chi}{\nu}\op{n}\lpar{\op{b}^{\dagger} + \op{b}} + \frac{\chi^{2}}{\nu^{2}}\op{n}^{2}}
    \lpar{e^{-\frac{\mi k}{\gamma}\lspar{\nu\op{N} + \omega\op{n} - \frac{\chi^{2}}{\nu}\op{n}^{2}}}\op{D}\lpar{\frac{\chi \op{n}}{\nu}}}\ket{\psi(0)}\\
    &=\bra{\psi(0)}\left(\op{N}+\frac{\chi}{\nu}\op{n}\op{b}^{\dagger}\lpar{1 - e^{\frac{\mi k}{\gamma}\nu}}\right.+\frac{\chi}{\nu}\op{n}\op{b}\lpar{1 - e^{-\frac{\mi k}{\gamma}\nu}}+\left.4\frac{\chi^{2}}{\nu^{2}}\op{n}^{2}\sin\lpar{\frac{k}{2\gamma}\nu}^{2}\right)\ket{\psi(0)}.
\end{aligned}
\end{equation}

For an initial wavefunction as the product of coherent states in the field and mirror, $\ket{\psi(0)} = \ket{\alpha}_{f}\otimes\ket{\beta}_{m}$, we have
\begin{equation}
\begin{aligned}
    &_{m}\bra{\beta}_{f}\bra{\alpha}\left(\op{N}+\frac{\chi}{\nu}\op{n}\op{b}^{\dagger}\lpar{1 - e^{\frac{\mi k}{\gamma}\nu}}+\frac{\chi}{\nu}\op{n}\op{b}\lpar{1 - e^{-\frac{\mi k}{\gamma}\nu}}\right.\left.4\frac{\chi^{2}}{\nu^{2}}\op{n}^{2}\sin\lpar{\frac{k}{2\gamma}\nu}^{2}\right)\ket{\alpha}_{f}\ket{\beta}_{m} =\\
    &|\beta|^{2} + \frac{\chi}{\nu}|\alpha|^{2}\beta^{*}\lpar{1-e^{\frac{\mi k\nu}{\gamma}}} + \frac{\chi}{\nu}|\alpha|^{2}\beta\lpar{1-e^{\frac{-\mi k\nu}{\gamma}}}+ 4\frac{\chi^{2}}{\nu^{2}}|\alpha|^{2}\lpar{1+|\alpha|^{2}}\sin\lpar{\frac{1}{2}\frac{k\nu}{\gamma}}^{2} = \bra{\psi_{k}}\op{N}\ket{\psi_{k}}
\end{aligned}
\end{equation}

\begin{figure}[htbp]
    \centering
    \includegraphics[width = \columnwidth]{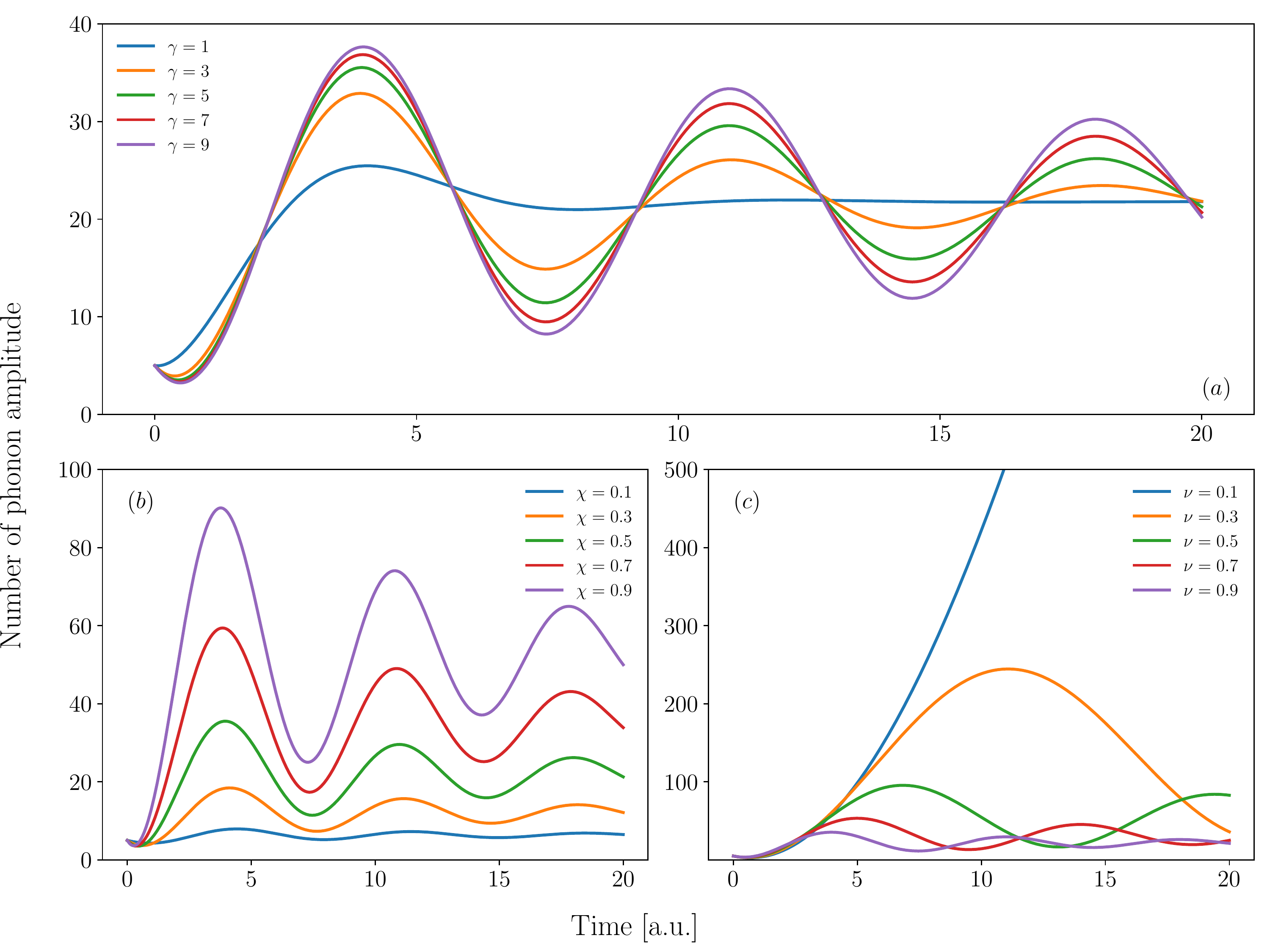}
    \caption{Number of modes $\braket{\op{N}}$ in the mirror. We variate the parameters involved in the decoherence evolution. a) Decaying parameter $\gamma$ between $1$ and $9$, maintaining $\chi = 0.5$ and $\nu = 0.9$. b) Coupling parameter $\chi$ between $0.1$ and $0.9$, maintaining $\gamma = 5$ and $\nu = 0.9$. c) Mirror oscillation frequency $\nu$ between $0.1$ and $0.9$, maintaining $\gamma = 5$ and $\chi = 0.5$. The initial conditions in the wavefunctions are $\ket{\alpha, \beta} = \ket{2\mi, 1 + 2\mi}$.}
    \label{fig:n_mirror}
\end{figure}

Therefore, the expectation value for the modes in the mirror is then
\begin{equation}
\begin{aligned}
    \braket{\op{N}} &= e^{-\gamma t}\sum\limits_{k = 0}^{\infty}\frac{\lpar{\gamma t}^{k}}{k!}\bra{\psi_{k}}\op{N}\ket{\psi_{k}}\\
    & = e^{-\gamma t}\sum\limits_{k = 0}^{\infty}\frac{\lpar{\gamma t}^{k}}{k!}\left(|\beta|^{2} + \frac{\chi}{\nu}|\alpha|^{2}\beta^{*}\lpar{1-e^{\frac{\mi k\nu}{\gamma}}}\right.+\left.\frac{\chi}{\nu}|\alpha|^{2}\beta\lpar{1-e^{\frac{-\mi k\nu}{\gamma}}} + 4\frac{\chi^{2}}{\nu^{2}}|\alpha|^{2}\lpar{1+|\alpha|^{2}}\sin\lpar{\frac{1}{2}\frac{k\nu}{\gamma}}^{2}\right)\\
    & = e^{-\gamma t}\left[|\beta|^{2}e^{\gamma t} + \frac{\chi}{\nu}|\alpha|^{2}\beta^{*}\lpar{e^{\gamma t} - e^{\gamma t e^{\frac{\mi\nu}{\gamma}}}}+\frac{\chi}{\nu}|\alpha|^{2}\beta\lpar{e^{\gamma t} - e^{\gamma t e^{\frac{-\mi\nu}{\gamma}}}} + \frac{\chi^{2}}{\nu^{2}}|\alpha|^{2}\lpar{1+|\alpha|^{2}}\right.\\
    &\qquad\times\left.\lpar{2e^{\gamma t} - \lpar{e^{\gamma t e^{\frac{\mi\nu}{\gamma}}} + e^{\gamma t e^{-\frac{\mi\nu}{\gamma}}}}}\right.,
\end{aligned}
\end{equation}
that returns the initial value at $t=0$, $\braket{\op{N}(t = 0)} = |\beta|^{2}$. For an initial condition $\ket{\psi(0)} = \ket{2\mi,1 + 2\mi}$, in Figure \ref{fig:n_mirror} we show how the number of phonon modes variates, modulates, and decays over time. The parameters that change are the decaying rate $\gamma$ (above), the coupling strength between mirror and field $\chi$ (middle), and the frequency of the mirror $\nu$ (below). The number of phonon modes is strictly positive, so we are seeing how different set of parameters involved leads to decaying dynamics in a more fast or slow fashion. There's a relevant feature of the dynamics when we observe the evolution over the set of decoherence parameters $\gamma$. Despite each figure showing the creation of phonons from the initial value of $\bar{\op{N}}(0) = 5$, when we look at the figure above, it is clear that there's a central value around $\bar{\op{N}}(t) \approx 22$ where other curves deviate when we increase or decrease the value of $\gamma$.

\subsubsection{Position quadrature $\braket{\op{b}^{\dagger} + \op{b}}$ in the mirror}
Turning the attention to the position variation of the mirror, we calculate the expectation value of twice the quadrature operator, taking the action of the wavefunction $\ket{\psi_{k}}$ over the operator we have
\begin{equation}
\begin{aligned}
    \bra{\psi_{k}}\op{b}^{\dagger} &+ \op{b}\ket{\psi_{k}} = \bra{\psi(0)}\lspar{\op{b}^{\dagger}e^{\frac{\mi k \nu}{\gamma}} + \op{b}e^{-\frac{\mi k \nu}{\gamma}} - 4\frac{\chi}{\nu}\op{n}\sin\lpar{\frac{k\nu}{2\gamma}}^{2}}\ket{\psi(0)},
\end{aligned}
\end{equation}
that for an initial wavefunction as the product of coherent states in the field and mirror, $\ket{\psi(0)} = \ket{\alpha}_{f}\otimes\ket{\beta}_{m}$, the expression reduces to
\begin{equation}
\begin{aligned}
    \bra{\psi_{k}}\op{b}^{\dagger} + \op{b}\ket{\psi_{k}} &= _{m}\bra{\beta}_{f}\bra{\alpha}\lspar{\op{b}^{\dagger}e^{\frac{\mi k \nu}{\gamma}} + \op{b}e^{-\frac{\mi k \nu}{\gamma}} - 4\frac{\chi}{\nu}\op{n}\sin\lpar{\frac{k\nu}{2\gamma}}^{2}}\ket{\alpha}_{f}\ket{\beta}_{m}\\ &= \beta^{*}e^{\frac{\mi k \nu}{\gamma}} + \beta e^{-\frac{\mi k \nu}{\gamma}} - 4\frac{\chi}{\nu}\left|\alpha\right|^{2}\sin\lpar{\frac{k\nu}{2\gamma}}^{2}
\end{aligned}
\end{equation}

\begin{figure}[htbp]
    \centering
    \includegraphics[width = \columnwidth]{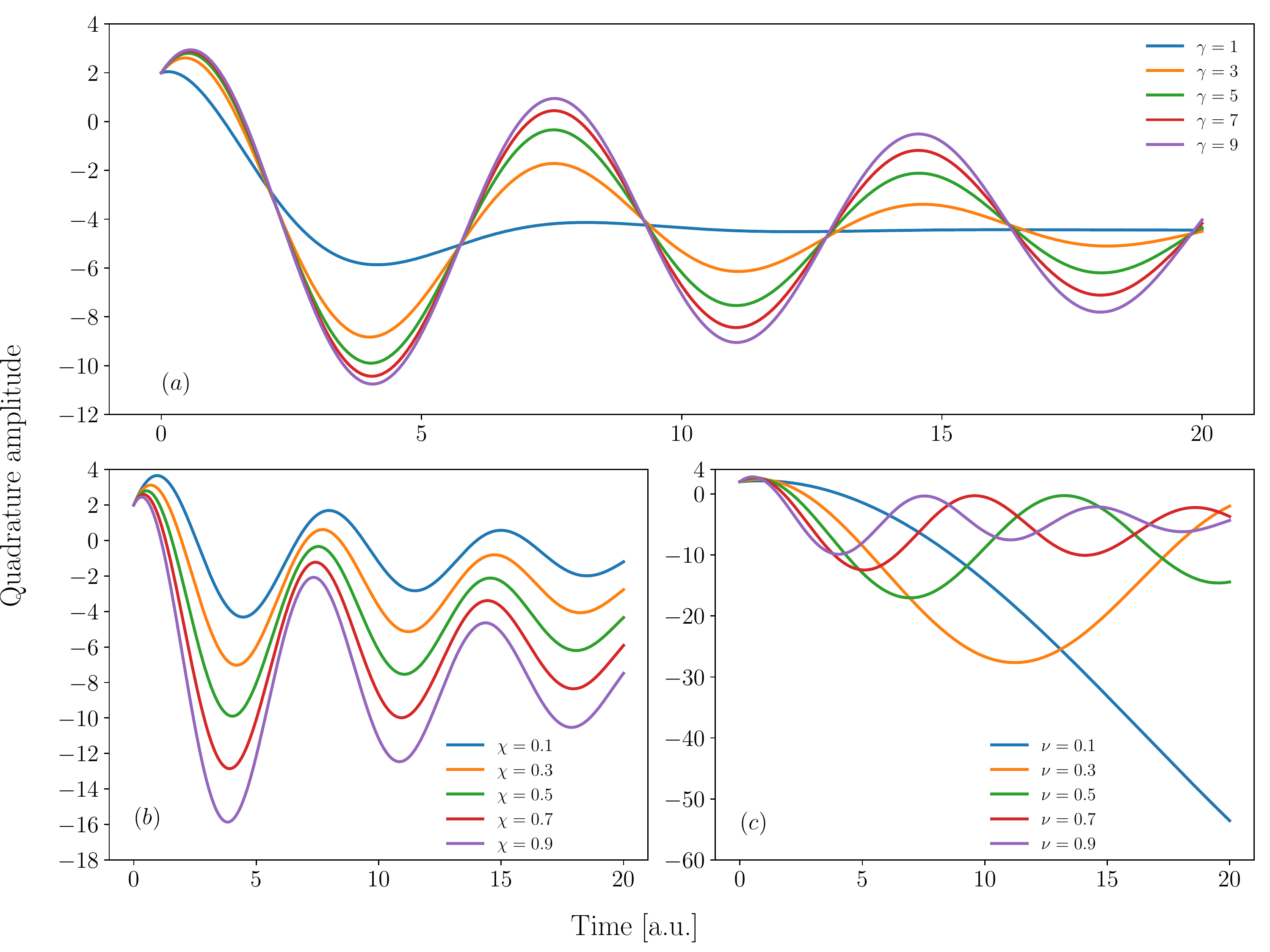}
    \caption{Position quadrature $\braket{\op{b}^{\dagger} + \op{b}}$ in the mirror. We variate the parameters involved in the decoherence evolution. a) Decaying parameter $\gamma$ between $1$ and $9$, maintaining $\chi = 0.5$ and $\nu = 0.9$. b) Coupling parameter $\chi$ between $0.1$ and $0.9$, maintaining $\gamma = 5$ and $\nu = 0.9$. c) Mirror oscillation frequency $\nu$ between $0.1$ and $0.9$, maintaining $\gamma = 5$ and $\chi = 0.5$. The initial conditions in the wavefunctions are $\ket{\alpha, \beta} = \ket{2\mi, 1 + 2\mi}$.}
    \label{fig:quad_mirror}
\end{figure}

This renders the expectation value for the position quadrature of the mirror as
\begin{equation}
\begin{aligned}
    \braket{\op{b}^{\dagger} + \op{b}} &= e^{-\gamma t}\sum\limits_{k = 0}^{\infty}\frac{\lpar{\gamma t}^{k}}{k!}\bra{\psi_{k}}\op{b}^{\dagger} + \op{b}\ket{\psi_{k}}\\
    &= e^{-\gamma t}\left(\beta^{*}e^{\gamma t e^{\frac{\mi\nu}{\gamma}}} + \beta e^{\gamma t e^{-\frac{\mi\nu}{\gamma}}}\right. - \left.\frac{\chi}{\nu}\left|\alpha\right|^{2}\lpar{2e^{\gamma t} - \lspar{e^{\gamma t e^{\frac{\mi\nu}{\gamma}}} + e^{\gamma t e^{-\frac{\mi\nu}{\gamma}}}}}\right),
\end{aligned}
\end{equation}
where the initial value at $t=0$ is $\braket{\lpar{\op{b}^{\dagger} + \op{b}}} = \beta^{*} + \beta \equiv 2\mathrm{Re}\lpar{\beta}$. In Figure \ref{fig:quad_mirror} we show how the position quadrature of the mirror variates, modulates, and decays over time. The parameters that change are the decaying rate $\gamma$ (above), the coupling strength between mirror and field $\chi$ (middle), and the frequency of the mirror $\nu$ (below). If we look at the figure above, where the decaying rate variates, we see again a central value $\approx 5$ where the curves change their amplitude according to the increase or decreasing value of $\gamma$.

\subsubsection{Number of photons $\braket{\op{n}}$ in the field}
As we expect from the behavior of the original system \eqref{ansol}, the expectation value for the number of photons in the field is a constant with a value of
\begin{equation}
\braket{\op{n}} = \left|\alpha\right|^{2},
\end{equation}
because every displacement operator and their exponential map involved commutes with the number operator $\op{n}$.

\subsection{Hong-ou-Mandel parameter and covariance}
We now take place to quantify and analyze a couple of statistical descriptors of the system, the Hong-ou-Mandel parameter, and the covariance between the photon and phonon modes of the field and the mirror, respectively.

The covariance of two well-behaved operators $\op{A}$ and $\op{B}$ is defined as
\begin{equation}
    \mathrm{cov}\lpar{\op{A}, \op{B}} := \braket{\op{A}\op{B}} - \braket{\op{A}}\braket{\op{B}},
\end{equation}
where $\braket{\cdot}$ is the standard expected value of an operator on a wavefunction. Note that the variance is just the autocovariance in the sense $\mathrm{var}\lpar{\op{A}} \equiv \mathrm{cov}\lpar{\op{A}, \op{A}} = \braket{\op{A}^{2}} - \braket{\op{A}}^{2}$. With this, we can define the Hong-ou-Mandel parameter as
\begin{equation}
    \op{O}_{\mathrm{HMp}} := \frac{\mathrm{var}\lpar{\op{O}}}{\braket{\op{O}}}.
\end{equation}

For our particular case, the expectation values are defined in terms of the action of the wavefunctions $\ket{\psi_{k}}$ on the designed operator $\op{A}$ and $\op{B}$ as
\begin{equation}
    \braket{\op{A}\op{B}} := e^{-\gamma t}\sum\limits_{k = 0}^{\infty}\;\frac{\lpar{\gamma t}^{k}}{k!}\braket{\psi_{k}\vert\op{A}\op{B}\vert\psi_{k}},\qquad \braket{\op{A}}^{2} := \lpar{e^{-\gamma t}\sum\limits_{k = 0}^{\infty}\;\frac{\lpar{\gamma t}^{k}}{k!}\braket{\psi_{k}\vert\op{A}\vert\psi_{k}}}^{2},
\end{equation}

\paragraph{Variance of $\op{n}$.} As stated before, the expectation value of the number of photons in the field is constant in time, then the variance evolution is effectively zero. Meaning that the Hong-ou-Mandel parameter has nothing to say about the correlation between the field itself.

\begin{figure}[ht]
    \centering
    \includegraphics[width = \textwidth]{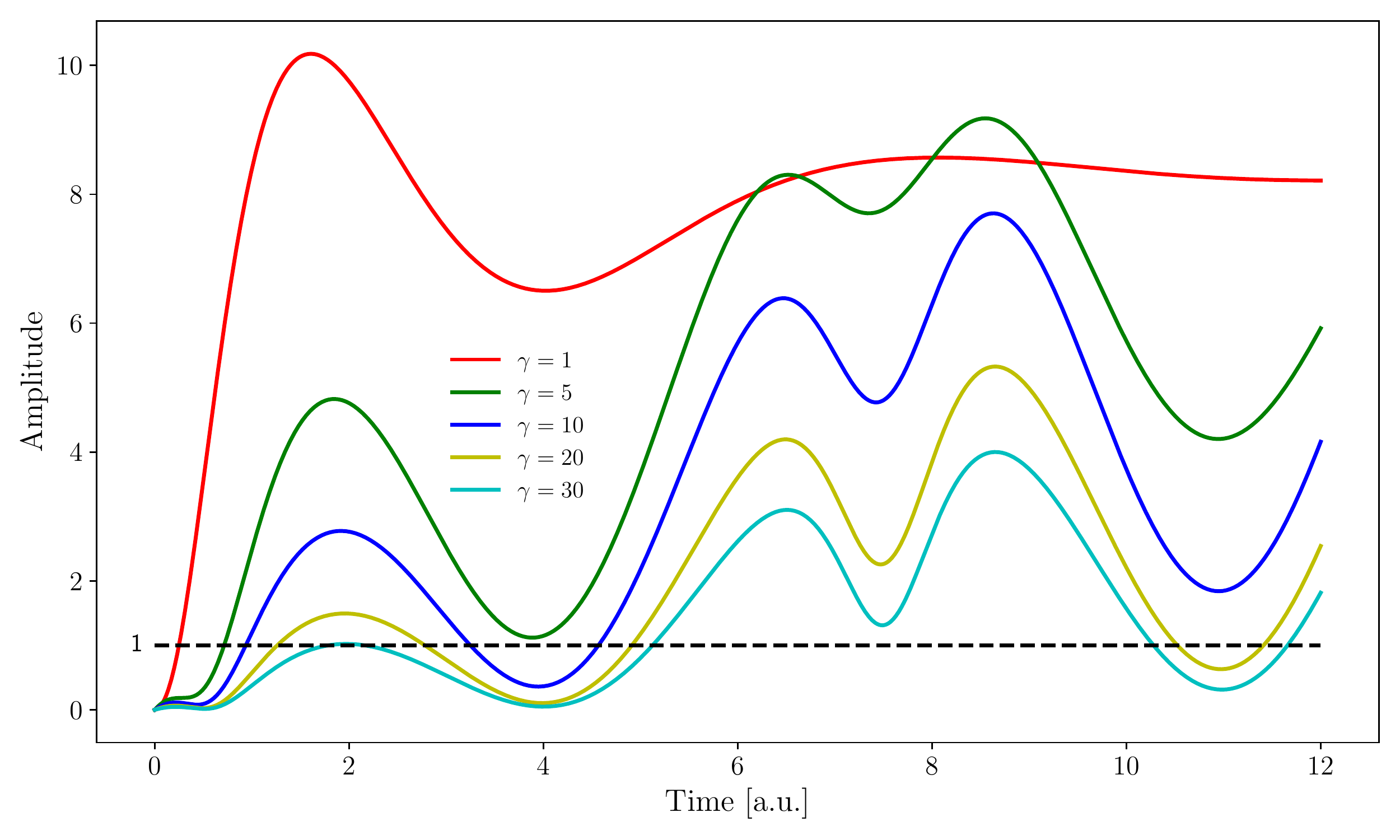}
    \caption{Hong-ou-Mandel parameter of the phonon mode number $\op{N}$, eq. \eqref{fig:hm_nb}. According to the definition, when the parameter goes above the unitary amplitude, we expect classical behavior in the evolution of the system. It is clear that this feature is directly proportional to the value of the decoherence rate $\gamma$. We set the values $\nu = 0.5$, $\chi = 0.9$, $\ket{\psi_{0}} = \ket{2\mi, 1 + 2\mi}$.}
    \label{fig:hm_nb}
\end{figure}

\paragraph{Variance of $\op{N}$.} The value of the variance for the phonon modes of the mirror is defined by $\braket{N^{2}} - \braket{N}^{2}$, with
\begin{equation}\label{N2}
\begin{aligned}
    \braket{\op{N}^{2}} &= e^{-\gamma t}\sum\limits_{k = 0}^{\infty}\frac{\lpar{\gamma t}^{k}}{k!} \lspar{|\beta|^{2} + \frac{\chi}{\nu}|\alpha|^{2}\beta^{*}\lpar{1-e^{\frac{\mi k\nu}{\gamma}}} + \frac{\chi}{\nu}|\alpha|^{2}\beta\lpar{1-e^{\frac{-\mi k\nu}{\gamma}}}+ 4\frac{\chi^{2}}{\nu^{2}}|\alpha|^{2}\lpar{1+|\alpha|^{2}}\sin\lpar{\frac{1}{2}\frac{k\nu}{\gamma}}^{2}}^{2}\\
    \braket{\op{N}}^{2} &= \lspar{e^{-\gamma t}\sum\limits_{k = 0}^{\infty}\frac{\lpar{\gamma t}^{k}}{k!} \lpar{|\beta|^{2} + \frac{\chi}{\nu}|\alpha|^{2}\beta^{*}\lpar{1-e^{\frac{\mi k\nu}{\gamma}}} + \frac{\chi}{\nu}|\alpha|^{2}\beta\lpar{1-e^{\frac{-\mi k\nu}{\gamma}}}+ 4\frac{\chi^{2}}{\nu^{2}}|\alpha|^{2}\lpar{1+|\alpha|^{2}}\sin\lpar{\frac{1}{2}\frac{k\nu}{\gamma}}^{2}}}^{2},
\end{aligned}
\end{equation}
where the sums in \eqref{N2} can be done analytically, giving cumbersome and lengthy expressions. Having these equations, we can calculate the Hong-ou-Mandel parameter for the number of phonons evolution in the mirror
\begin{equation}
    \op{N}_{\mathrm{HMp}} = \frac{\braket{\op{N}^{2}} - \braket{\op{N}}^{2}}{\braket{\op{N}}},
\end{equation}
as shown in the Figure \ref{fig:hm_nb}. We can see that the transition from classical to non-classical behavior, and vice versa, is directly proportional to the amplitude of the decoherence rate $\gamma$. 

\paragraph{Covariance between the photon and phonon modes.} We can quantify how the evolution of the phonon modes of the mirror is correlated to the (constant) evolution of the photon modes of the field. For this, we take the covariance 
\begin{equation}
    \mathrm{cov}\lpar{\op{n}, \op{N}} = \braket{\op{n}\op{N}} - \braket{\op{n}}\braket{\op{N}},
\end{equation}
where the joint expect value $\braket{\op{n}\op{N}}$ is
\begin{equation}
\begin{aligned}
    \braket{\op{n}\op{N}} &= e^{-\gamma t}\sum\limits_{k = 0}^{\infty}\frac{\lpar{\gamma t}^{k}}{k!}\bra{\psi_{k}}\op{n}\op{N}\ket{\psi_{k}}\\
    &= |\alpha|^{2}|\beta|^{2} + \frac{\chi}{\nu}|\alpha|^{2}\lpar{1 + |\alpha|^{2}}\beta^{*}\lpar{1-e^{\frac{\mi k\nu}{\gamma}}} + \frac{\chi}{\nu}|\alpha|^{2}\lpar{1 + |\alpha|^{2}}\beta\lpar{1-e^{\frac{-\mi k\nu}{\gamma}}}\\
    &+ 4\frac{\chi^{2}}{\nu^{2}}|\alpha|^{2}\lpar{1 + 3|\alpha|^{2} + |\alpha|^{2}}\sin\lpar{\frac{1}{2}\frac{k\nu}{\gamma}}^{2},
\end{aligned}
\end{equation}
that differs from $\braket{\op{n}}\braket{\op{N}}$ due the emergence of a $\op{n}^{3}$ term. Despite this, it seems that there are no new functions involving time, just a scale in the amplitude given by the photon modes $\op{n}^{k}$.

\subsection{Husimi $\op{Q}$ function}
The Husimi $\op{Q}$ function is defined as the expectation value of the density matrix on the coherent basis as \cite{}
\begin{equation}\label{qhusimi}
    \mathrm{Q}(\epsilon) = \frac{1}{\pi}\bra{\epsilon}\op{\rho}\ket{\epsilon},
\end{equation}
where $\ket{\epsilon}$ is the coherent state ket on a certain suitable basis. When we are dealing with multiple subsystems, the Husimi function can be taken as the pseudo-distribution on their respective phase space of the subsystem. If we have a system decomposed as $\op{H} = \op{H}_{f}\otimes\op{H}_{m}$, the Husimi function is then
\begin{equation}\label{Qhusimi}
    \mathrm{Q}(\epsilon, \zeta) = \frac{1}{\pi} ~_{m}\bra{\zeta}_{f}\bra{\epsilon}\op{\rho}\ket{\epsilon}_{f}\ket{\zeta}_{m},
\end{equation}
where $f$ and $m$ are the subscripts for \emph{field} and \emph{mirror} subsystem, respectively.

We know from \eqref{rhok} and \eqref{psik} that the density matrix of the field-mirror system is defined as
\begin{equation}
    \op{\rho} = e^{-\gamma t}\sum\limits_{k = 0}^{\infty} \frac{\lpar{\gamma t}^{k}}{k!} \ket{\psi_{k}}\bra{\psi_{k}},
\end{equation}
where the $k$-th component of the wavefunction $\psi_{k}$ is
\begin{equation}
    \ket{\psi_{k}} = \op{D}^{\dagger}\lpar{\frac{\chi}{\nu} \op{n}} e^{-\frac{\mi k}{\gamma}\lpar{\omega\op{n} + \nu\op{N} - \frac{\chi^{2}}{\nu^{2}}\op{n}^{2}}} \op{D}^{\dagger}\lpar{\frac{\chi}{\nu} \op{n}} \ket{\psi(0)},
\end{equation}
for some initial wavefunction $\ket{\psi(0)}$. 
When $\ket{\psi(0)} = \ket{\alpha}_{f}\ket{\beta}_{m}$, the Husimi function \eqref{Qhusimi} can be obtained in the coherent basis $\ket{a}_{f}\ket{b}_{m}$ as
\begin{equation}
    \mathrm{Q}\lpar{a, \alpha; b, \beta} := \frac{1}{\pi}\bra{b}\bra{a}\op{\rho}\ket{a}\ket{b},
\end{equation}
that renders explicitly the terms
\begin{equation}
    \bra{b}\bra{a}\op{\rho}\ket{a}\ket{b} \equiv \bra{b}\bra{a}\psi_{k}\rangle\langle\psi_{k}\ket{a}\ket{b}.
\end{equation}

The explicit calculation shows that
\begin{equation}\label{sandwichQ}
\begin{aligned}
    \bra{b}\bra{a}\psi_{k}\rangle &= \bra{b}\bra{a}\op{D}^{\dagger}\lpar{\frac{\chi}{\nu} \op{n}} e^{-\frac{\mi k}{\gamma}\lpar{\omega\op{n} + \nu\op{N} - \frac{\chi^{2}}{\nu^{2}}\op{n}^{2}}} \op{D}\lpar{\frac{\chi}{\nu} \op{n}} \ket{\alpha}\ket{\beta}\\
    &= \bra{b}\op{D}^{\dagger}\lpar{\frac{\chi}{\nu} \vert a\vert^{2}} e^{-\frac{\mi k}{\gamma}\lpar{\omega a^{*}\alpha + \nu\op{N} - \frac{\chi^{2}}{\nu^{2}}(a^{*})^{2}(\alpha)^{2}}} \op{D}\lpar{\frac{\chi}{\nu} \vert \alpha\vert^{2}} \ket{\beta}\\
    &= \bra{b + \tfrac{\chi}{\nu} \vert a\vert^{2}} e^{-\frac{\mi k}{\gamma}\lpar{\omega a^{*}\alpha + \nu\op{N} - \frac{\chi^{2}}{\nu^{2}}(a^{*})^{2}(\alpha)^{2}}} \ket{\beta + \tfrac{\chi}{\nu} \vert \alpha\vert^{2}}\\
    &= e^{-\frac{\mi k}{\gamma}\lspar{\omega a^{*}\alpha + \nu\lpar{b + \tfrac{\chi}{\nu} \vert a\vert^{2}}^{*}\lpar{\beta + \tfrac{\chi}{\nu} \vert \alpha\vert^{2}} - \frac{\chi^{2}}{\nu^{2}}(a^{*})^{2}(\alpha)^{2}}} \braket{a|\alpha}\braket{b + \tfrac{\chi}{\nu} \vert a\vert^{2}|\beta + \tfrac{\chi}{\nu} \vert \alpha\vert^{2}},
\end{aligned}
\end{equation}\
for $a, \alpha, b, \beta \in \mathbb{C}$. From \eqref{sandwichQ} we can obtain the dagger $\langle\psi_{k}\ket{a}\ket{b}$, that finally gives expression
\begin{equation}\label{qf}
    \bra{b}\bra{a}\psi_{k}\rangle\langle\psi_{k}\ket{a}\ket{b} = \frac{1}{\pi} e^{-\frac{\mi k}{\gamma}\lspar{\omega f(a, \alpha) + \nu g(a, \alpha; b, \beta) - \frac{\chi^{2}}{\nu^{2}} h(a, \alpha)}} e^{-\vert a - \alpha\vert^{2}} e^{-\left| \lpar{b - \beta} + \tfrac{\chi}{\nu}\lpar{\vert a\vert^{2} - \vert \alpha\vert^{2}}\right|^{2}},
\end{equation}
where the functions inside the first exponential are defined as
\begin{equation}
\begin{aligned}
    &f(a, \alpha) = a^{*}\alpha - a\alpha^{*}\\
    &g(a, \alpha; b, \beta) = \lpar{b + \tfrac{\chi}{\nu} \vert a\vert^{2}}^{*}\lpar{\beta + \tfrac{\chi}{\nu} \vert \alpha\vert^{2}} - \lpar{b + \tfrac{\chi}{\nu} \vert a\vert^{2}}\lpar{\beta + \tfrac{\chi}{\nu} \vert \alpha\vert^{2}}^{*}\\
    &h(a, \alpha) = (a^{*})^{2}(\alpha)^{2} - (a)^{2}(\alpha^{*})^{2}.
\end{aligned}
\end{equation}

Finally, with the equations \eqref{rhot} and \eqref{qf}, the Husimi function is explicitly given by
\begin{equation}\label{Qfunc}
     \mathrm{Q}\lpar{a, \alpha; b, \beta} = \frac{1}{\pi}e^{-\gamma t}e^{\gamma t e^{-\frac{\mi}{\gamma}\lspar{\omega f(a, \alpha) + \nu g(a, \alpha; b, \beta) - \frac{\chi^{2}}{\nu^{2}} h(a, \alpha)}}} e^{-\vert a - \alpha\vert^{2}} e^{-\left| \lpar{b - \beta} + \tfrac{\chi}{\nu}\lpar{\vert a\vert^{2} - \vert \alpha\vert^{2}}\right|^{2}},
\end{equation}
that resembles Gaussians in phase space displaced by $\alpha$ and $\beta$, in which amplitude is modulated by the first terms that are time-dependent.

\begin{figure}[ht]
    \centering
    \includegraphics[width = \columnwidth]{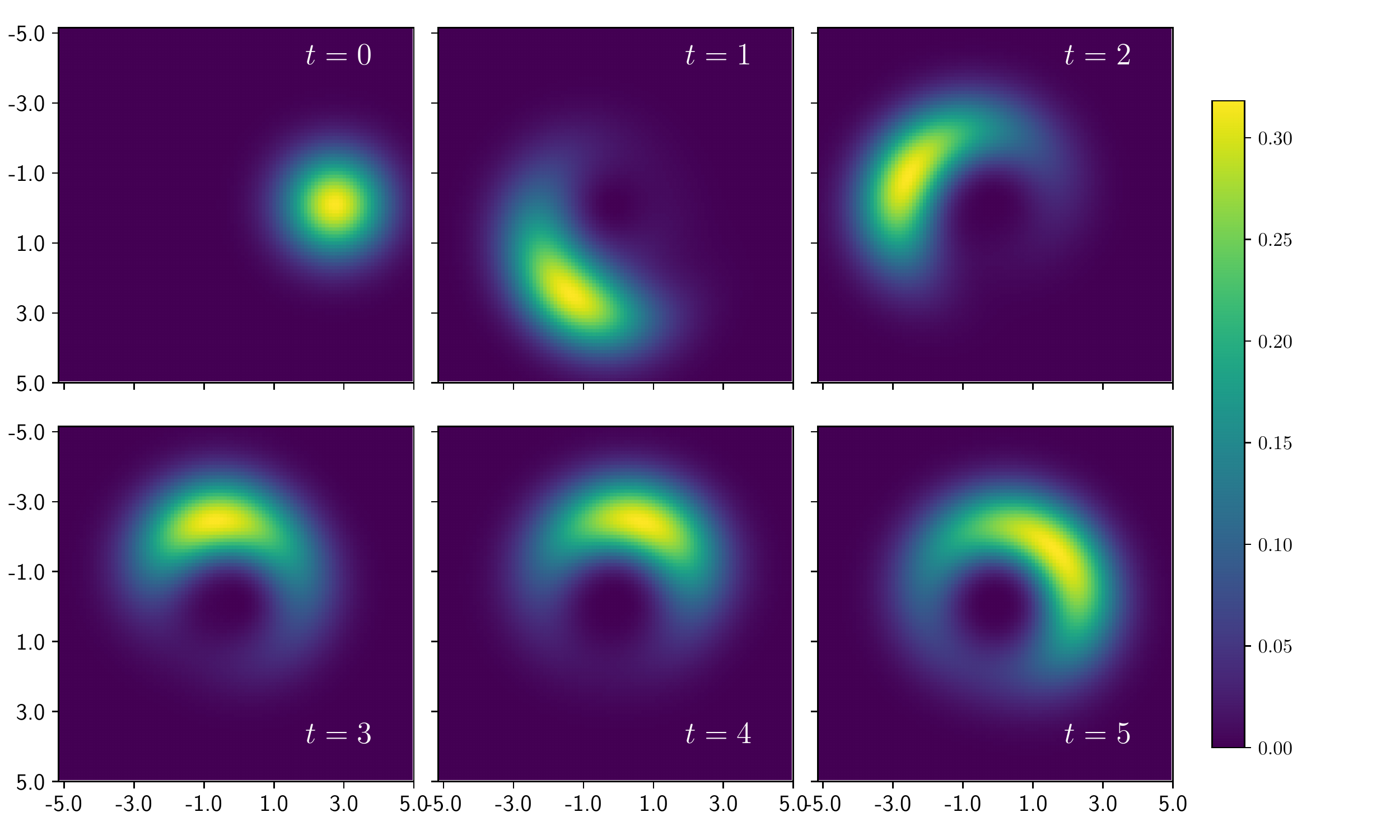}
    \caption{Husimi $\op{Q}$ function for the field. We set $\ket{\psi_{0}} = \ket{2\mi, 1 + 2\mi}$, $\nu = 0.5$, $\chi = 0.9$ and $\gamma = 20$}
    \label{fig:qalpha}
\end{figure}
\begin{figure}[ht]
    \centering
    \includegraphics[width = \columnwidth]{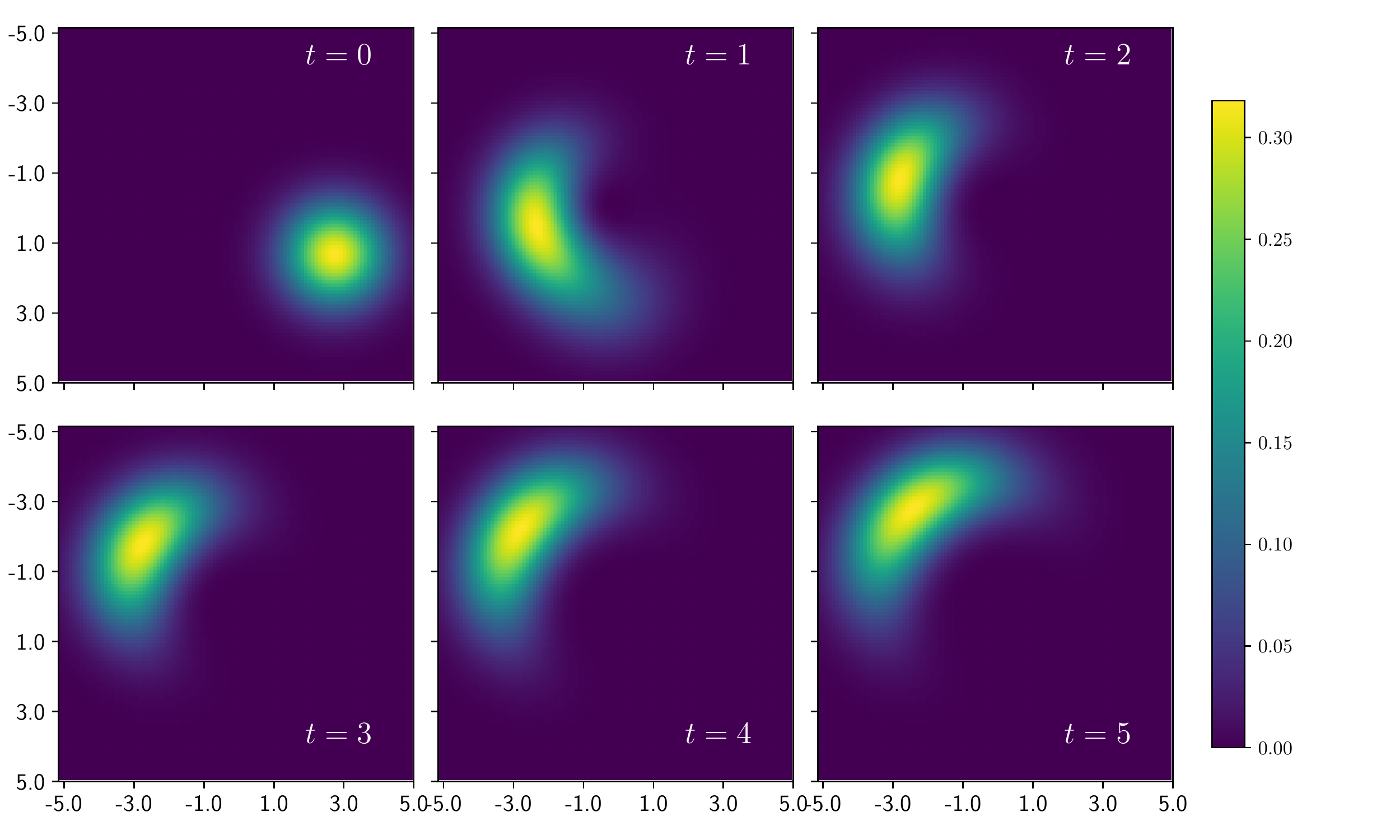}
    \caption{Husimi $\op{Q}$ function for the mirror. We set $\ket{\psi_{0}} = \ket{2\mi, 1 + 2\mi}$, $\nu = 0.5$, $\chi = 0.9$ and $\gamma = 20$}
    \label{fig:qbeta}
\end{figure}

In Figures \ref{fig:qbeta} and \ref{fig:qalpha} we show the evolution of the initial states of the field and the mirror, respectively. We set $\ket{\psi_{0}} = \ket{2\mi, 1 + 2\mi}$, $\nu = 0.5$, $\chi = 0.9$ and $\gamma = 20$. According to the expected value for the number of photons in the field, we see that the $\op{Q}$-function evolves as a harmonic oscillator; on the other side, the $\op{Q}$-function for the mirror evolves according from what the Hong-ou-Mandel parameters suggest: at some point, the initial coherent state transit from a non-classical to classical description, when $t > 4$ we see that the decoherence stall the movement around phase-space. 

\section{Analysis and conclusions}
We solve the mirror-field interaction using the intrinsic decoherence scheme from the complete Milburn equation \eqref{rhot}. For this particular driven system, a \emph{decoherence parameter} $\gamma$ defines the strength in the decaying dynamics. We obtain the significant expected values for the number of mechanical phonon modes, and the position quadrature; also we observe the constant nature of the number of photon modes inside the cavity. Although the radiation pressure is parametrically coupled to the harmonic motion of the mirror, the number of quanta remains the same. For a better understanding of the feedback between the quantized field and the mechanical mirror, we calculate the covariance between the number of photons and phonons, giving just an amplitude scaling proportional to $\op{n}^{k}$, but no new time-dependent dynamics. The Hong-ou-Mandel parameter is obtained for the photon and phonon modes. The photon modes have nothing new to say, but the phonon modes give insights into how the decoherence made the initial coherent state in the mirror transit from a non-classical to a classical description. This last can be seen also in the Husimi $\op{Q}$-functions we show for the field and the mirror, where the behavior on phase-space enables us to talk about the quantized nature of the photon modes and the macroscopic nature of the moving mirror.

\section*{Declarations}
\subsection*{Funding} 
Consejo Nacional de Ciencia y Tecnolog\'ia (Postdoctoral Grant 2021 and 2022) 

\subsection*{Acknowledgements}
A.R. Urz\'ua thanks Dr. Francisco R\'ecamier (ICF UNAM) for the help in understanding the Hong-ou-Mandel parameter. A.R. also thanks ICF UNAM for the logistical support during the postdoctoral stay.

\printbibliography
\end{document}